\documentclass{PoS}
\PoS{PoS(LAT2005)080}

\usepackage{amssymb}

\usepackage{epsfig}

\usepackage{graphicx}
\usepackage[figuresright]{rotating}

\newcommand{\AmS}{{\protect\the\textfont2
  A\kern-.1667em\lower.5ex\hbox{M}\kern-.125emS}}

\title{Light meson masses and decay constants in 2+1 flavour domain wall QCD}

\ShortTitle{Light meson masses and decay constants in 2+1 flavour domain wall QCD}

\author{D.J.~Antonio, K.C.~Bowler, P.A.~Boyle, M.A.~Clark, B.~Jo\'o, A.D.~Kennedy, R.D.~Kenway, C.M.~Maynard and \speaker{ R.J~Tweedie }\\
	School of Physics,\\
        The University of Edinburgh,\\
	Edinburgh EH9 3JZ, UK\\
	E-mail: \email{s0459477@sms.ed.ac.uk},\email{ kcb@ph.ed.ac.uk},\email{ paboyle@ph.ed.ac.uk},\\
	\email{ mike@ph.ed.ac.uk},\email{ bj@ph.ed.ac.uk},\email{ adk@ph.ed.ac.uk},\\
        \email{ r.d.kenway@ed.ac.uk},\email{ cmaynard@ph.ed.ac.uk},\email{ rjt@ph.ed.ac.uk}}

\author{A.~Yamaguchi\\
	Department of Physics and Astronomy,\\
	The University of Glasgow,\\
	University Avenue,\\
	Glasgow G12 8QQ, UK\\ 
	\email{a.yamaguchi@physics.gla.ac.uk}}

\author{RBC and UKQCD Collaborations}

\abstract{
We present results for light meson masses and psedoscalar meson decay
constants in 2+1 flavour domain wall QCD with the DBW2 and Iwasaki
gauge actions, using lattices with linear sizes in the range 1.6 to
2.2fm and $u$ and $d$ quark masses as low as one quarter of the
strange quark mass. All data were generated on the QCDOC machines at
the University of Edinburgh and Brookhaven National
Laboratory. Despite large residual masses and a limited number of sea
quark mass values with which to perform chiral extrapolations, our
results agree with experiment and scale within errors.  }

\FullConference{XXIIIrd International Symposium on Lattice Field Theory\\
                 25-30 July 2005\\
                 Trinity College, Dublin, Ireland}
     
\begin{document}

\section{INTRODUCTION}
The application of the RHMC algorithm~\cite{Clark:2004cp} to lattice
QCD with Ginsparg-Wilson quarks together with the computational power
of QCDOC has resulted in the ability to generate, and perform
measurements on, dynamical 2+1 flavour configurations for the first
time. However, before embarking on a large-scale and (potentially)
computationally costly production run, it is necessary to explore the
available parameter space with (several) smaller and, therefore,
computationally cheaper initial trials. This work focuses on the
recent ensembles generated on the QCDOC machines for exactly this
purpose.

\section{SIMULATION PARAMETERS}
The analysis was carried out on 2+1 flavour domain wall fermion
configurations generated on the QCDOC machines. The standard domain
wall Dirac operator~\cite{Furman:1994ky} and Pauli-Villars field with
the action introduced in~\cite{Vranas:1997da} was used.  The gauge
action was defined by
\begin{equation}
  S_G[U] = - \frac{\beta}{3}
          \left[(1-8\,c_1) \sum_{x;\mu<\nu} P[U]_{x,\mu\nu}
        + c_1 \sum_{x;\mu\neq\nu} R[U]_{x,\mu\nu}\right]
\end{equation}
where $P[U]_{x,\mu\nu}$ and $R[U]_{x,\mu\nu}$ represent the real part
of the trace of the path ordered product of links around the $1\times
1$ plaquette and $1\times 2$ rectangle, respectively, in the $\mu,\nu$
plane at the point $x$, and $\beta \equiv 6/g^2$ with $g$ the bare
coupling constant.  For the DBW2 gauge
action~\cite{Takaishi:1996xj,deForcrand:1999bi}, the coefficient $c_1$
was chosen to be $-1.4069$, whereas for
Iwasaki~\cite{Iwasaki:1983ck,Iwasaki:1984cj} $c_{1} = -0.331$. In the
DBW2 case the ensembles have three different $\beta$ values: 0.72,
0.764 and 0.78, while for the Iwasaki case they have two: 2.13 and
2.2. All ensembles were generated using the RHMC
algorithm~\cite{Clark:2004cp} with a trajectory length of 0.5, volume
of $16^{3}$x32, fifth dimension length of 8, and a domain wall height
of 1.8. Two different ensembles were generated at each $\beta$ value,
one with a light isodoublet with mass $m_{ud}=\frac{1}{2}m_{s}$ and
one where the light sea quark masses are both equal to our rough
estimate of the strange quark mass, $m_{s}$. An additional DBW2
ensemble with a light isodoublet mass $m_{ud}=\frac{1}{4}m_{s}$ at
$\beta$=0.72 was generated. The number of trajectories in each
ensemble and the number of configurations used in the analysis are
shown in table~\ref{tab:datasets}. The ensembles had $am_{ud}$ = 0.01,
0.02 or 0.04 and $am_{s}$ = 0.04.

Measurements were made with up to four valence quark masses in the
range 0.01 to 0.04 on each ensemble and, on some of the ensembles,
correlators were measured with sources on multiple time planes to
improve our statistics. Several types of smearing were used, in
particular, point sources, wall sources, and hydrogen-like
wavefunction smearing where one or two of the quark propagators in a
meson correlator were smeared at the source. In general, simultaneous
fits to local and smeared correlators were performed throughout. This
analysis aggregates to more than thirty thousand trajectories and more
than one hundred and twenty thousand measurements.

The integrated auto-correlation time for the pseudoscalar meson
correlator was calculated on several of these ensembles and found to
be of order 100 trajectories. The correlators were oversampled and
averaged into bins of size between five and ten depending on whether
the separation between measurements was five or ten trajectories. Our
observed errors stabilised with bins of this size in good agreement
with the calculated integrated auto-correlation time. A full
correlated analysis was then performed with the binned data as input.

\begin{table}
\begin{center}
\begin{tabular}{|c|c|c|c|c|c|c|}
\hline
Action&$\beta$&$\frac{m_{ud}}{m_{s}}$&$am_{ud}$&$N_{traj}$&$N_{meas}$&$am_{\rm res}$\\\hline\hline
DBW2&0.72&1.0&0.04&3395&475&\\
DBW2&0.72&0.5&0.02&6000&1000&0.0106(1)\\
DBW2&0.72&0.25&0.01&2340&269&\\
DBW2&0.764&1.0&0.04&1615&150&\\
DBW2&0.764&0.5&0.02&1800&88&\raisebox{1.5ex}[0pt]{0.00546(7)}\\
DBW2&0.78&1.0&0.04&1620&113&\\
DBW2&0.78&0.5&0.02&1505&202&\raisebox{1.5ex}[0pt]{0.00437(6)}\\
\hline
Iwasaki&2.13&1.0&0.04&2380&139&\\
Iwasaki&2.13&0.5&0.02&2450&146&\raisebox{1.5ex}[0pt]{0.0104(2)}\\
Iwasaki&2.2&1.0&0.04&4565&713&\\
Iwasaki&2.2&0.5&0.02&3175&435&\raisebox{1.5ex}[0pt]{0.0065(1)}\\
\hline
\end{tabular}
\end{center}
\caption{\label{tab:datasets}\label{tab:res_mass} RHMC 2+1 flavour datasets and the value of the residual mass in the chiral limit.}
\end{table}

\section{RESULTS}
\subsection{The Residual Mass}
The residual mass is a measure of the violation of chiral
symmetry~\cite{Furman:1994ky}. In our ensembles the length of the
fifth dimension is relatively short, hence there is a significant
left-right coupling between the quark fields on opposite walls.  The
residual mass was calculated from the ratio of the point-split
pseudoscalar density $J_{5}$ at the middle of the fifth dimension to
the pseudoscalar density P built from the fields on the
walls~\cite{Furman:1994ky}
\begin{equation}
am_{\rm res} = \frac{ \Sigma_{\vec{x},\vec{y}} \langle
J_5(\vec{y},t)P(\vec{x},0)\rangle }{\Sigma_{\vec{x},\vec{y}} \langle
P(\vec{y},t) P(\vec{x},0)\rangle}.
\end{equation}

\begin{figure}
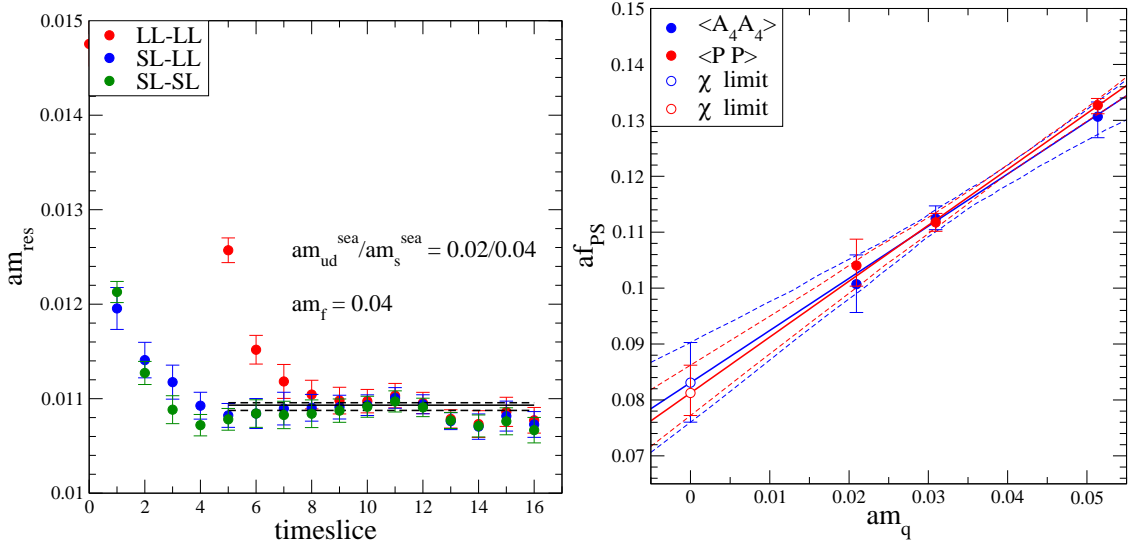

\epsfig{file=mres_2.13_2+1f_eff_mass.eps, width=.49\textwidth}
\epsfig{file=fpi_AA_PP.eps, width=.49\textwidth}
\caption{\label{plot:mres_eff_mass} {\bf LEFT:} Residual mass for $\beta$ = 2.13 ensemble with $m_{ud}=\frac{1}{2}m_{s}$. Different colours correspond to whether zero, one or two quark propagators are smeared at the source. \label{plot:f_pi}{\bf RIGHT:} Pseudoscalar meson decay constant using two different methods versus quark mass for the DBW2 $\beta$=0.72 ensembles.} 
\end{figure}

A good signal for the value of $am_{\rm res}$ was observed, as can be
seen in figure~\ref{plot:mres_eff_mass} (left). 

Chiral extrapolations were performed by defining the quark mass
\begin{equation}
am_{q} \equiv a(m_{f} + m_{\rm res}(m_{f})),
\end{equation}
where $am_{f}$ is the valence quark mass and $m_{\rm res}(m_{f})$ is
the residual mass measured using quark propagators with that quark
mass, and taking $am_{q} \rightarrow 0$ using only the points where
the valence quark mass is equal to the $u$ and $d$ quark masses in the
sea. Other than in the DBW2 $\beta$=0.72 case, where a linear fit to
three points was performed, straight lines were drawn through the two
available points. The values of the residual mass obtained in the
chiral limit are shown in table~\ref{tab:res_mass}. These
values correspond to a residual mass of $\sim$5-30 MeV.

\subsection{Light hadron masses}
Pseudoscalar and vector meson masses are extracted by performing
simultaneous double cosh fits to extract both the ground and first
excited states. 

Chiral extrapolations were performed in an analogous way to the
residual mass by extrapolating the results to $m_{q}$ = 0. In the
case of the pseudoscalar mass this takes the form
\begin{equation}
\label{eqn:PSchiral}
(am_{PS})^{2} = B(am_{q}^{val=sea}) + A.
\end{equation}
Given the low statistics, an acceptable slight deviation of $A$ from
zero, typically less than 2$\sigma$, is observed. The lattice spacing
was obtained from $m_{\rho}$ in the chiral limit by performing a
linear chiral extrapolation of the vector meson mass
\begin{equation}
am_{V} = C(am_{q}^{val=sea}) + D. 
\end{equation}
This, together with the physical kaon mass and substituting
$am_{q}^{val=sea} \rightarrow am_{q_{1}}+am_{q_{2}}$ in
eq.(~\ref{eqn:PSchiral}), allowed the evaluation of the strange quark
mass by setting $am_{q_{1}}=am_{ud}=0$ and $am_{q_{2}}=am_{s}$.


%
%

\subsection{Pseudoscalar meson decay constants}
The pseudoscalar meson decay constant defined by 
\begin{equation}\label{fPS}
af_{PS} = \frac{Z_{A}\langle 0 | A_{4} | P \rangle }{am_{PS}}
\end{equation}
was calculated in two ways. 
In the first method $Z_{A}$ was obtained from the axial Ward-Takahashi
identity	
\begin{equation}
Z_{A}\langle\partial_{\mu}A_{\mu} \mathcal{O} \rangle = 2(am_{q}+am_{\rm
res})\langle P \mathcal{O} \rangle
\end{equation}
which can be expressed in terms of the pseudoscalar meson correlator,
$C_{PP}$, and the pseudoscalar meson axial correlator,
$C_{PA_{4}}$. $\langle0|A_{4}|P\rangle$ cancels when $Z_{A}$ is
substituted in eq.(\ref{fPS}) and hence only $C_{PP}$ and the value of
the residual mass are required in order to evaluate $f_{PS}$
\begin{equation}
af_{PS} = \frac{2 (am_{f} + am_{\rm res})\langle 0 | P | P
\rangle}{(am_{PS})^{2}}.
\end{equation}
The red symbols in figure \ref{plot:f_pi} (right) show the chiral
extrapolation of $f_{PS}$ for the DBW2 $\beta$=0.72 ensembles.

In the second method, we calculated the value of $Z_{A}$ explicitly
from a ratio designed to remove $\mathcal{O}$(a) and suppress
$\mathcal{O}$($a^{2}$) lattice artifacts~\cite{Blum:2000kn}.
\begin{equation}\label{Z_A}
Z_{A} = \frac{C(t+\frac{1}{2})+C(t-\frac{1}{2})}{4L(t)} +
\frac{C(t+\frac{1}{2})}{(L(t)+L(t+1))}
\end{equation}
where $C(t)$ and $L(t)$ are the correlators of the pseudoscalar
density with the partially conserved and local axial currents
respectively. A simultaneous fit to both point-point and
wall/smeared-point correlators is used to extract this ratio. The
axial-axial correlator was then used in combination with the value of
$Z_{A}$ to calculate the pseudoscalar meson decay constant. The chiral
extrapolation of $f_{PS}$ using this method is shown by the blue
symbols in figure~\ref{plot:f_pi} (right). We see good agreement
between both methods.

Using the lattice spacing and the strange quark mass calculated from
the vector meson and pseudoscalar meson chiral extrapolations, we are
able to calculate $f_{K}$ and extract the ratio
$\frac{f_{K}}{f_{\pi}}$.

%
%

\subsection{Scaling}
\begin{figure}
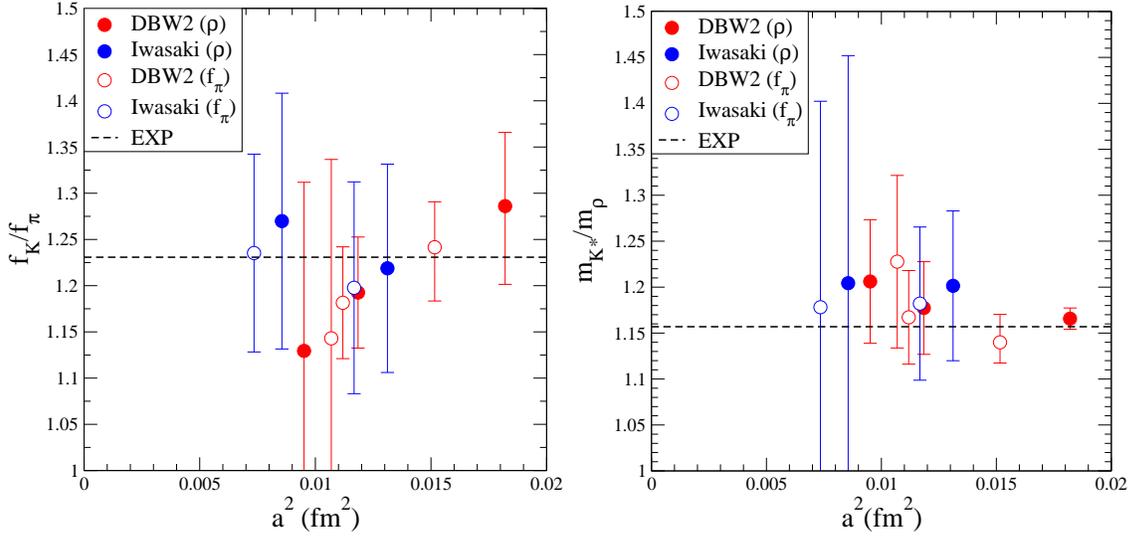

\epsfig{file=lattice_05_fK_fpi_scaling.eps, width=.49\textwidth}
\epsfig{file=lattice_05_mKstar_amrho_scaling.eps, width=.49\textwidth} 
\caption{\label{plot:scaling1}\label{plot:scaling2}Scaling behaviour of {\bf LEFT:} $\frac{f_{K}}{f_{\pi}}$ and {\bf RIGHT:} $\frac{m_{K^{*}}}{m_{\rho}}$ for all ensembles. Different colours correspond to different gauge actions. Open and closed symbols correspond to setting the lattice spacing from $f_{\pi}$ or $m_{\rho}$ in the chiral limit respectively.}
\end{figure}

The ensembles generated for this analysis with several different
lattice spacings and two different gauge actions have
$\mathcal{O}$($a^{2}$) discretisation
errors. Figure~\ref{plot:scaling1} (left) shows the ratio
$\frac{f_{K}}{f_{\pi}}$ plotted against lattice spacing squared. The
different colour symbols correspond to the different gauge actions,
Iwasaki in blue and DBW2 in red. We have set the lattice spacing in
two different ways. The closed symbols correspond to setting the
lattice spacing from the $\rho$ meson mass, while the open symbols are
from setting the lattice spacing from $f_{\pi}$. Note that the errors
are still quite large. The furthest two points to the right are the
DBW2 $\beta$=0.72 points where we have been able to perform a chiral
fit compared to drawing straight lines through the points so their
errors are better estimated. Even with the large and crude error
estimate for the other results, it may be concluded that there is an
indication of scaling and, if anything, scaling may slightly better
for the Iwasaki gauge action than for DBW2. Figure~\ref{plot:scaling2}
(right) shows a similar plot to figure~\ref{plot:scaling1} (left), for
$\frac{m_{K*}}{m_{\rho}}$ and similar behaviour is seen.

\section{SUMMARY}
Several ensembles have been generated on the QCDOC machines using the
domain wall fermion formulation with two different gauge actions,
several $\beta$ values and multiple sea quark masses. These ensembles
have a relatively small volume and limited statistics as they were
primarily generated to search parameter space for a larger production
run. A fifth dimension size of eight produces a residual mass larger
than would be acceptable for such a production run. Even with these
drawbacks it is still possible to calculate the light hadron spectrum
and pseudoscalar decay constants obtaining results which are
consistent with experiment and scale within large errors. At present
we are running an ensemble with a larger volume of $24^{3}$x$64$ and a
fifth dimension size of 16 using the Iwasaki gauge action at $\beta =
2.13$.

\section{ACKNOWLEDGEMENTS}
We thank Sam Li and Meifeng Lin for help generating the datasets used
in this work. We thank Dong Chen, Norman Christ, Saul Cohen, Calin
Cristian, Zhihua Dong, Alan Gara, Andrew Jackson, Chulwoo Jung,
Changhoan Kim, Ludmila Levkova, Xiaodong Liao, Guofeng Liu, Robert
Mawhinney, Shigemi Ohta, Konstantin Petrov and Tilo Wettig for
developing with us the QCDOC machine and its software. This
development and the resulting computer equipment used in this
calculation were funded by the U.S. DOE grant DE-FG02-92ER40699, PPARC
JIF grant PPA/J/S/1998/00756 and by RIKEN. This work was supported by
PPARC grant PPA/G/O/2002/00465.


\vspace{-0.5cm}
\providecommand{\href}[2]{#2}\begingroup\raggedright\endgroup



\end{document}